\documentclass[twocolumn,prl,showpacs,amsmath,amssymb]{revtex4}

\usepackage{graphicx}

\def\be{\begin{equation}}
\def\ee{\end{equation}}
\def\bea{\begin{eqnarray}}
\def\eea{\end{eqnarray}}
\def\bma{\begin{mathletters}}
\def\ema{\end{mathletters}}

\def\0{\overline{0}}

\def\q0{\underline{0}}

\def\H{{\cal H}}

\def\L{{\cal L}}

\def\tr{\mbox{tr}}
\def\one{\leavevmode\hbox{\small1\normalsize\kern-.33em1}}

\def\bra#1{\langle#1|} \def\ket#1{|#1\rangle}
\def\braket#1#2{\langle#1|#2\rangle}

\def\proj#1{\ket{#1}\!\bra{#1}}

\begin{document}

\title{ Key distillation from Gaussian states by Gaussian operations
 }

\author{M. Navascu\'es$^{1}$, J. Bae$^{1}$, J. I. Cirac$^2$, M. Lewenstein$^3$, A. Sanpera$^3$ and A. Ac\'\i n$^{1}$}

\affiliation{ $^1$Institut de Ci\`encies Fot\`oniques, Jordi
Girona 29, Edifici Nexus II, E-08034 Barcelona, Spain\\
$^2$Max-Planck Institut f\"ur Quantenoptik, Hans-Kopfermann Str.
1, D-85748 Garching, Germany\\
$^3$ Institut f\"ur Theoretische Physik, Universit\"at Hannover,
D-30167 Hannover, Germany }

\date{\today}

%%%%%%%%%%%% Abstract %%%%%%%%%%%%%%%%%%%%%%%%%%%

\begin{abstract}

We study the secrecy properties of Gaussian states under Gaussian
operations. Although such operations are useless for quantum
distillation, we prove that it is possible to distill a secret key
secure against any attack from sufficiently entangled Gaussian
states with non-positive partial transposition. Moreover, all such
states allow for key distillation, when Eve is assumed to perform
finite-size coherent attacks before the reconciliation process.

\end{abstract}


\maketitle

%\section{Introduction}
%\label{intro}

Quantum Information Theory (QIT) analyzes the possibilities
offered by quantum states for information encoding and
transmission. New processes such as teleportation \cite{telep},
more powerful algorithms \cite{shor} or completely secure data
transmission \cite{BB84} become possible by exploiting quantum
effects. Most of the QIT applications use entanglement as
resource. It is therefore one of the most important tasks of QIT
to determine if a given state is entangled, and if yes, whether
its entanglement is ``useful" for information processing.
Distillable states \cite{distil,qpa} posses such useful
entaglement: using several copies of them and local operations and
classical communication (LOCC), one can create a smaller number of
maximally entangled states, which can be used for variety of
quantum information tasks. Given a state, its distillable
entanglement, $E_D$, measures the amount of pure-state
entanglement that can be extracted from it. Particularly important
for cryptography is another measure of entanglement $K_D$: it
specifies the number of secret bits that can be extracted from a
quantum state using LOCC. Obviously, $K_D\ge E_D$, since a secret
key can always be extracted from distilled maximally entangled
states, using, for instance, the Ekert protocol \cite{Ekert}.
Entangled states which cannot be distilled, i.e. for which
$E_D=0$, exist and are termed ``bound entangled" \cite{bound}.
Although it was unclear whether these states could contain useful
entanglement, it has been recently shown that there exist bound
entangled states for which $K_D>E_D=0$ \cite{HO}.

All of the above results were originally considered for finite
dimensional systems. More recently, many of these concepts have
been translated to the infinite dimensional case \cite{BP}. In
these systems, a key role is played by the set of Gaussian states
and Gaussian operations. First of all, it naturally appears in
experiments. In fact, non-Gaussian operations turn out to be very
challenging from an experimental point of view. Moreover, the
theoretical analysis of Gaussian states and operations is also
simplified since all their properties can be expressed in terms of
finite-dimensional matrices.

To analyze the limitations and possibilities offered by the
Gaussian scenario is a relevant issue. The teleportation of
coherent states of light have been experimentally demonstrated
\cite{exptel}. Quantum cryptography has also been successfully
translated into the Gaussian regime. Gottesman and Preskill
proposed to use squeezed states and homodyne measurements in a
``prepare \& measure'' scheme \cite{cryptogaus}. Actually, no
squeezing is required, since coherent states are already
sufficient for a secure key distribution with Gaussian operations
\cite{GG,IVC}. The experimental implementation of a coherent-state
protocol has been recently realized in \cite{nature}. From a more
fundamental point of view, it is interesting to know which
information tasks can be achieved in the Gaussian regime. It is
known that all Gaussian states have positive Wigner function, i.e.
there is a local variable model reproducing all the (symmetrically
ordered) correlations for Gaussian states. Despite the existence
of this local description, these states and operations are useful
for teleporting quantum states, or for secure information
transmission, applications that are impossible in a classical
scenario.

An important  negative result in this context was obtained  in
Refs. \cite{GC,ESP}: quantum distillation by Gaussian operations
is impossible. That is, although it is known that all Gaussian
states with non-positive partial transposition (NPPT) are
distillable \cite{GDCZ}, any distillation protocol must include a
non-Gaussian operation. This can be rephrased saying that {\sl all
entangled mixed states are bound entangled in the Gaussian
scenario}. In this scenario, it is natural to define the
corresponding Gaussian versions of entanglement measures, that
specify the entanglement properties of these states under Gaussian
Local Operations and Classical Communication (GLOCC). For example,
$GE_D$ defines the Gaussian Distillable Entanglement. The results
of \cite{GC,ESP} imply that $GE_D=0$. However, these states may
still be useful, since perhaps {\sl secret bits} can be extracted
from them using GLOCC.

%The ability to distribute secrecy is one of the distinctive
%features of entanglement: any entangled state can be mapped into a
%probability distribution containing secret correlations \cite{AG}.
In a similar way as for distillable entanglement, for any
entangled state one can denote by $GK_D$ the amount of secret bits
that can extracted by GLOCC. This quantity refers to the
distribution of a secret key, and in principle  is independent of
entanglement distillability. Thus, secret bits ($GK_D$) appear as
a resource (measure) especially suitable for the analysis of
entangled Gaussian states under Gaussian operations: while
$GE_D=0$ for all states, there are states with positive $GK_D$
\cite{cryptogaus,IVC}.

In this work we analyze the secrecy properties of Gaussian states
under Gaussian operations \cite{eisert}. We consider a protocol
where Alice and Bob measure their shared entangled state at the
single-copy level and process the obtained classical results
\cite{assump}. This type of protocols can easily be translated to
prepare \& measure schemes. First, we study the security of our
protocol when Eve applies an incoherent attack, that is she
individually measures her state before any key extraction process
is started. Under these conditions, all NPPT states turn out to be
secure. The same conclusion holds when Eve applies a coherent
attack on a finite number of symbols. All these cases would
correspond to a situation where Eve's quantum memory has a finite
coherence time. Then, we consider more powerful attacks and show
that our security proof ceases to work for some NPPT states.
Finally, using the recent results of Ref. \cite{Renner}, we
demonstrate the security of sufficiently entangled Gaussian states
against any attack.

We consider quantum systems of $n$ canonical degrees of freedom,
often called modes, $\H=\L^2(\Re^n)$. The commutation relations
for the canonical coordinates $R=(X_1,P_1,\ldots,X_n,P_n)=
(R_1\ldots,R_{2n})$ read $[R_a,R_b]=i(J_n)_{ab}$, where
$a,b=1,\ldots,2n$ and
\begin{equation}
    J_n=\oplus_{i=1}^n J \quad\quad J\equiv \begin{pmatrix}
    0 & 1 \cr
    -1 & 0
  \end{pmatrix} .
\end{equation}
The characteristic function, $\chi_\rho(x)$, of a state $\rho$ is
defined as $\chi_\rho(x)\equiv\tr(\rho W(x))$, where
$W(x)=\exp(-ix^TR)$ are the so-called Weyl operators. Gaussian
states are those states such that $\chi_\rho$ is a Gaussian
function,
\begin{equation}
\label{chfunction}
    \chi_\rho(x)=\exp(ix^Td-\frac{1}{4}x^T\gamma x) .
\end{equation}
where $d$ is a $2n$ real vector, called displacement vector (DV),
and $\gamma$ is a $2n\times 2n$ symmetric real matrix, known as
covariance matrix (CM). The positivity condition of $\rho$ implies
that $\gamma-iJ_n\geq 0$. All the information about $d$ and
$\gamma$ is contained in the first and second moments $\tr(\rho
R_i)$ and $\tr(\rho R_iR_j)$.

In what follows we consider two parties, Alice and Bob, that share
a state $\rho$ in a composite systems of $n+m$ modes. The global
CM is
\begin{equation}
\label{gammaab}
    \gamma_{AB}=\begin{pmatrix}\gamma_A & C \cr
    C^T & \gamma_B \end{pmatrix}\geq i J_{n+m} ,
\end{equation}
where $\gamma_A$ ($\gamma_B$) is the CM for the $n$-mode
($m$-mode) Gaussian state of system $A$ ($B$). The entanglement
properties of $\rho$ are completely specified by its CM.

The effect of partial transposition at the level of CMs can be
understood from the fact that this map is equivalent to
time-reversal. After partial transposition on, say, $A$, the sign
of Alice's momenta is changed while the rest of canonical
coordinates is kept unchanged. Denote by $\theta$ the matrix equal
to the identity for the position coordinates and minus the
identity for the momenta. Partial transposition means that
$\gamma_{AB}\rightarrow\gamma'_{AB}=\theta_A\gamma_{AB}\theta_A$.
Therefore, the state $\rho$ has positive partial transposition
(PPT) when $\gamma'_{AB}$ defines a positive operator, that is
$\gamma'_{AB}\geq iJ_{n+m}$. The PPT criterion provides a
necessary and sufficient condition for separability for $1\times
1$ \cite{11mode} and $1\times N$ Gaussian states \cite{WW}, while
it is only a necessary condition for the rest of systems
\cite{WW}. As said above, the non-positivity of the partial
transposition is a necessary and sufficient condition for
distillability \cite{GDCZ}.

Two known results will play an important role in what follows.
First, any NPPT Gaussian state of $n+m$ modes can be mapped by
GLOCC into an NPPT $1\times 1$ Gaussian and symmetric state
\cite{GDCZ}, whose CM, see Eq. (\ref{gammaab}), is
\begin{equation}
\label{symmst}
    \gamma_A=\gamma_B=\begin{pmatrix}
    \lambda & 0 \cr
    0 & \lambda
  \end{pmatrix} \quad\quad
  C=\begin{pmatrix}
    c_x & 0 \cr
    0 & -c_p
  \end{pmatrix}
\end{equation}
where $\lambda\geq 0$ and $c_x\geq c_p\geq 0$. The positivity
condition reads $\lambda^2-c_xc_p-1\geq \lambda(c_x-c_p)$ while
the entanglement (NPPT) condition gives
\begin{equation}
\label{entcond}
    \lambda^2+c_xc_p-1< \lambda(c_x+c_p) .
\end{equation}
And second, given an $n$-mode Gaussian state $\rho_1$ with CM
$\gamma_1$, it is always possible to construct a $2n$ mode pure
Gaussian state $\ket{\Psi_{12}}$ such that
$\tr_2(\proj{\Psi_{12}})=\rho_1$ \cite{schmidt}. The global CM
$\gamma_{12}$, see Eq. (\ref{gammaab}), has $\gamma_A=\gamma_1$
and
\begin{eqnarray}
\label{purif}
  \gamma_B&=&\theta\gamma_1\theta\nonumber\\
    C&=&J_n S\left(\oplus_{i=1}^{n}\sqrt{\lambda_k^2-1}
    \one_2\right)S^{-1}\theta ,
\end{eqnarray}
where $\{\lambda_k\}$ defines the symplectic spectrum of
$\gamma_1$ and $S$ is the symplectic matrix such that $S^T\gamma_1
S$ is diagonal.

Having collected all these facts, let us describe how Alice and
Bob can distill a key from a distillable $n\times m$ Gaussian
state using only Gaussian operations. Since all the NPPT Gaussian
states can be mapped into symmetric and entangled states of two
modes by GLOCC, we restrict our analysis to this type of states.
In equivalent terms, one can think that the first step in the key
distillation protocol is the GLOCC transformation of \cite{GDCZ}
that transforms any NPPT state into an entangled state of this
family. Then, both parties measure the $X$ quadrature, where $X_A$
and $X_B$ denote the measured operator and $x_A$ and $x_B$ the
obtained outcome. After communication, they only accept those
cases where $|x_A|=|x_B|=x_0$ \cite{notepr}. Each party associates
the logical bit 0 (1) to a positive (negative) result with the
probability $p\,(i,j)$, with  $i,j=0,1$. This process transforms
the quantum state into a list of correlated classical bits between
Alice and Bob. Their error probability, that is the probability
that their symbols do not coincide, is given by
\begin{equation}\label{errab}
    \epsilon_{AB}=\frac{\sum_{i\neq j}p\,(i,j)}
    {\sum_{i,
    j}p\,(i,j)}=\frac{1}{1+\exp
    \left(\frac{4c_xx_0^2}{\lambda^2-c_x^2}\right)} .
\end{equation}

In order to establish a key, Alice and Bob will now apply the
advantage distillation protocol introduced by Maurer
\cite{Maurer}. Alice generates the random bit $b$. Then, she
chooses $N$ items from her list of symbols, $\vec
b_A=(b_{A1},b_{A2},\ldots,b_{AN})$, and sends to Bob the vector
$\vec b$ such that $b_{Ai}+b_i=b \mbox{ mod }2,\,\forall\,
i=1,\ldots,N$, together with the list of chosen symbols. Bob
computes $b_{Bi}+b_i$ for his corresponding symbols, and if all
the results are equal, $b_{Bi}+b_i=b',\,\forall\, i$, the bit is
accepted. If not, the symbols are discarded and the
process is repeated for another vector. %Notice that the
%protocol can be very inefficient, since if $\epsilon_{AB}$ is
%large, only few vectors are accepted. However,
The new error probability is (see also \cite{AMG,bruss})
\begin{equation}\label{errABN}
    \epsilon_{AB,N}=\frac{(\epsilon_{AB})^N}{(1-\epsilon_{AB})^N+
    (\epsilon_{AB})^N}\leq\left(\frac{\epsilon_{AB}}
    {1-\epsilon_{AB}}\right)^N ,
\end{equation}
that tends to an equality for $N\rightarrow\infty$. %Note that in
%the limit of large $N$, only those cases where there have been no
%errors, $b=b'$, contribute to the distilled key.

What is the information that Eve can obtain? As usual, all the
environment, all the degrees of freedom outside Alice and Bob's
systems should be accessible to her. This means that the global
state including Eve is pure, $\ket{\Psi_{ABE}}$, and such that
$\tr_E(\proj{\Psi_{ABE}})=\rho_{AB}$ \cite{notepur}. Denote by
$\ket{e_{\pm\pm}}$ Eve's states when Alice and Bob have projected
onto $\ket{\pm x_0}$. For the case of individual attacks, it was
shown in \cite{AMG} that Eve's error in the estimation of the
final bit $b$ is bounded from below by a term proportional to
$|\braket{e_{++}} {e_{--}}|^N$. Therefore, Alice and Bob can
establish a key if (see \cite{AMG} for more details)
\begin{equation}
\label{keycond}
    \frac{\epsilon_{AB}}{1-\epsilon_{AB}}<|\braket{e_{++}}
    {e_{--}}| .
\end{equation}
More precisely, if this condition is fulfilled, there is always a
finite $N$ such that the new list of symbols can be distilled into
a secret key using one-way protocols \cite{CK}.

From Eq. (\ref{purif}), one can compute the global pure state
including Eve. Note that taking the Gaussian purification does not
imply any loss of generality on Eve's individual attack
\cite{notepur}. %Indeed, her attack can still be non-Gaussian,
%depending on the applied measurement.
After projecting on $\ket{\pm x_0}$, Eve has a Gaussian state of
two modes, with the
CM and DV for the states $\ket{e_{\pm\pm}}$ given by
\begin{eqnarray}
\label{evestate}
    d_{++}&=&-\frac{\sqrt{\lambda^2+\lambda(c_x-c_p)
    -c_xc_p-1}}{\lambda+c_x}\,(0,0,x_0,x_0)
    \nonumber\\
    \gamma_{++}&=&\begin{pmatrix}
    \gamma_{x} & 0 \cr
    0 & \gamma_{x}^{-1}
  \end{pmatrix} \quad\quad \gamma_{x}=\begin{pmatrix}
    \lambda & c_x \cr
    c_x & \lambda
  \end{pmatrix} ,
\end{eqnarray}
while $\gamma_{--}=\gamma_{++}$ and $d_{--}=-d_{++}$. Now, the
overlap between these two states is given by
\begin{equation}
\label{eveov}
    |\braket{e_{++}}{e_{--}}|^2=%\int\W_{++}\W_{--}=
    \exp\left(-\frac{4(\lambda^2+\lambda(c_x-c_p)
    -c_xc_p-1)x_0^2}{\lambda+c_x}\right) .
\end{equation}
Substituting Eqs. (\ref{errABN}) and (\ref{eveov}) in
(\ref{keycond}) one can check, after some algebra, that this
condition is equivalent to the entanglement condition of
(\ref{entcond}). That is, all the distillable (NPPT) Gaussian
states allow a secure key distribution under individual attacks
using Gaussian operations. Moreover, the limits for NPPT
entanglement and key distillation also coincide if Eve measures in
a coherent way a finite number $N_E\ll N$ of states {\sl before}
the reconciliation process \cite{note}. Interestingly, these
limits hold for any $x_0$, and measurements of arbitrary
resolution \cite{notepr}.

\begin{figure}
  % Requires \usepackage{graphicx}
  \includegraphics[width=7cm]{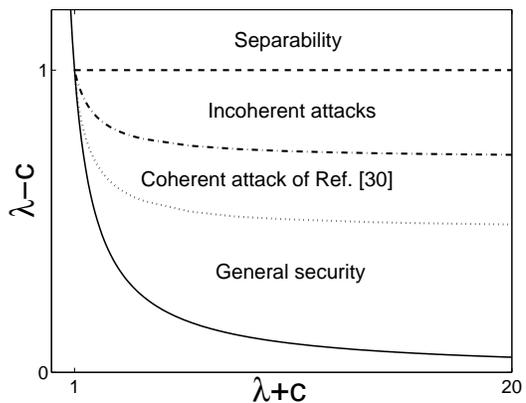}\\
  \caption{Security analysis of symmetric
  $1\times 1$ Gaussian states when $c_x=c_p=c$. All physical states
  are above the solid line. The dashed line defines the
  entanglement limit, that coincides with the security
  bound against incoherent attacks.
  States below the dotted line are secure against
  any attack. }\label{secan}
\end{figure}

One may wonder what happens when Eve is no longer restricted to
individual attacks. Indeed, a more powerful Eve could wait until
the end of the advantage distillation protocol and measure in a
coherent way all her $N$ symbols \cite{sing}. One can see that the
corresponding security condition is similar to Eq.
(\ref{keycond}), but replacing Eve's states overlap by its square.
This new inequality is violated by some NPPT states (see figure
\ref{secan}). Note that this only implies that the analyzed
protocol is not good for these states in this more general
scenario.

Nevertheless, using the recent techniques of Ref. \cite{Renner},
we can find states for which the presented protocol allows to
extract common bits secure against any attack. After a successful
$X$ measurement on the state (\ref{symmst}), Alice, Bob and Eve
share an effective state
\begin{eqnarray}
\label{psiabe22}
    \ket{\Psi_{ABE}^{2\times 2}}&=&\sqrt{\frac{1-\epsilon_{AB}}{2}}
    (\ket{++}\ket{e_{++}}+\ket{--}\ket{e_{--}})\nonumber\\
    &+&\sqrt{\frac{\epsilon_{AB}}{2}}(\ket{+-}\ket{e_{+-}}+
    \ket{-+}\ket{e_{-+}}) .
\end{eqnarray}
Then, it has been shown in Ref. \cite{Renner} that the amount of
secret bits, $R$, that Alice and Bob can extract from their known
quantum state \cite{assump} by means of protocols using one-way
communication is
\begin{equation}\label{rate}
    R\geq I(x_A:x_B)-S(\rho_{AB}^{2\times 2}) ,
\end{equation}
where $I$ is the mutual information between their measurement
outputs and $S(\rho_{AB}^{2\times 2})$ the von Neumann entropy of
their reduced state. In this case, this condition turns out to be
dependent on $x_0$. Thus, for any state, one has to look for a
value of $x_0$ such that $R>0$. Although we were not able to solve
this problem analytically, it can be attacked using numerical
methods. For example, the security curve for those states such
that $c_x=c_p$ (see Eq. (\ref{gammaab})) is shown in figure
\ref{secan}.

One can envisage different ways of improving the previous security
analysis, e.g. finding better measurements for Alice and Bob or
new ways of processing their measurement outcomes. A more
interesting possibility consists of allowing the honest parties to
manipulate in a coherent way several copies of their local states.
Actually, in the study of $GK_D$ with full generality, one should
deal with joint (although local and Gaussian) operations by Alice
and Bob. This defines a new type of {\sl Gaussian quantum privacy
amplification} protocols \cite{qpa} different from entanglement
distillability where Alice and Bob's goal is simply to factor Eve
out \cite{GC,eisert}.

A related open question that deserves further investigation is
whether secret bits can be extracted from PPT Gaussian states,
i.e. strict bound entangled states (cf. \cite{HO}). At present, we
know that our scheme does not work for any PPT state. The details
of the proof are quite involved and will be given elsewhere.

\begin{figure}
  % Requires \usepackage{graphicx}
  \includegraphics[width=8cm]{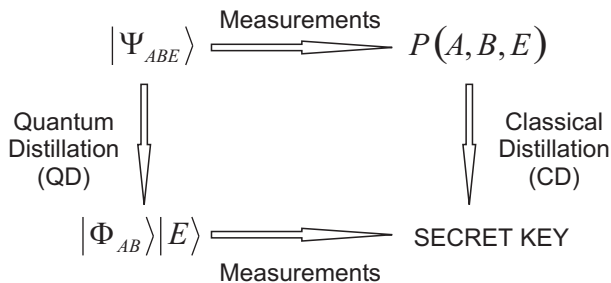}\\
  \caption{Schematic representation of
  quantum and classical key distillation techniques
  from quantum states. In this work, Alice and Bob
  only perform Gaussian operations.}\label{qcdist}
\end{figure}

Quantum and classical distillation (QD and CD) protocols are two
techniques that allow to extract secret bits from entangled
states. A schematic comparison between them is shown in figure
\ref{qcdist}. In finite systems, there are examples of
non-distillable quantum states for which the CD branch is possible
\cite{HO}. Moving to continuous variables systems and the Gaussian
scenario, QD techniques are useless for key-agreement
\cite{GC,ESP}. Our analysis proves that CD is still useful for (i)
all NPPT states under finite coherent attacks and (ii)
sufficiently entangled NPPT states under general attacks. Thus,
our results close the gap between NPPT entanglement and security
for the case of finite-size coherent attacks.

%To conclude, in this work the secrecy properties of Gaussian
%states under Gaussian operations are analyzed. We introduce a
%simple protocol consisting of single-copy measurements and
%classical processing of the obtained outcomes.

%\section{Acknowledgements}

We acknowledge discussions with D. Bru\ss, M. Christandl, J.
Eisert, F. Grosshans and M. Plenio. This work has been supported
by the Deutsche Forschungsgemeinschaft, the EU (projects RESQ and
QUPRODIS), the Kompetenzenznetzwerk
``Quanteninformationsverarbeitung", and the Generalitat de
Catalunya.


\begin{thebibliography}{99}
\bibitem{assump}
As for the entanglement distillability scenario, it is assumed
that Alice and Bob share a known Gaussian state.
\bibitem{notepr}
Although this only happens with zero probability, one can include
finite widths $dx_A=dx_B=dx$ for Alice and Bob's outcomes. For
small widths, all our conclusions hold by continuity and the
modified protocol has finite key rate. Interestingly, the security
claims against finite coherent attacks are also valid for
arbitrary large widths.
\bibitem{notepur}
The global state is specified by the local state $\rho_{AB}$, up
to a unitary operation on Eve's system. Thus, from Eve's point of
view, all the purifications are equivalent.
\bibitem{note}
Similarly as in Ref. \cite{AMG}, one can prove that Eq.
(\ref{keycond}) again defines the security condition for finite
coherent attacks.
\end{thebibliography}
\end{document}